\documentstyle[aps,twocolumn,epsf,floats]{revtex}

\newcommand{\la}[1]{\label{#1}}
\newcommand{\be}{\begin{equation}}
\newcommand{\ee}{\end{equation}}
\newcommand{\ba}{\begin{eqnarray}}
\newcommand{\ea}{\end{eqnarray}}
\newcommand{\bi}{\begin{itemize}}
\newcommand{\ei}{\end{itemize}}

\newcommand{\fig}{Fig.~}
\newcommand{\figs}{Figs.~}
\newcommand{\nr}[1]{(\ref{#1})}
\newcommand{\tr}{{\rm Tr\,}}

\newcommand{\fr}[2]{{\frac{#1}{#2}}}
\newcommand{\msbar}{\overline{\mbox{\rm MS}}}

\newcommand{\eq}{Eq.\,}
\newcommand{\eqs}{Eqs.\,}
\newcommand{\h}{{\hspace{0.5 cm}}}
\newcommand{\half}{\mbox{${1\over2}$}}

\newcommand{\etal}{{\em et al}}

\def\lsi{\raise0.3ex\hbox{$<$\kern-0.75em\raise-1.1ex\hbox{$\sim$}}}
\def\gsi{\raise0.3ex\hbox{$>$\kern-0.75em\raise-1.1ex\hbox{$\sim$}}}

\newcommand{\gsim}{\mathop{\gsi}}

\begin{document}
\twocolumn[\hsize\textwidth\columnwidth\hsize\csname
@twocolumnfalse\endcsname


\title{The universal properties of the electroweak phase transition$^*$}

\author{%
\hfill 
K. Kajantie,$\!^{\rm a,b}$ M. Laine,$\!^{\rm a,b}$
K. Rummukainen,$\!^{\rm c}$ M. Shaposhnikov$^{\rm a}$ 
and M. Tsypin$^{\rm d}$
\hfill\raisebox{21mm}[0mm][0mm]{\makebox[0mm][r]{NORDITA-98/69HE}}%
\raisebox{17mm}[0mm][0mm]{\makebox[0mm][r]{September 1998}}
}
\address{$^{\rm a}$Theory Division, CERN, CH-1211 Geneva 23, 
Switzerland}%

\address{$^{\rm b}$Department of Physics, P.O.Box 9,
00014 University of Helsinki, Finland}

\address{$^{\rm c}$Nordita, Blegdamsvej 17, DK-2100 Copenhagen \O,
Denmark}

\address{$^{\rm d}$Department of Theoretical Physics,
Lebedev Physical Institute, 117924 Moscow, Russia}

\date{\today}

\maketitle

\begin{abstract}
The first order electroweak phase transition in the Standard Model
turns into a regular cross-over at a critical Higgs mass $m_{H,c} \sim
75$\,GeV\@.  At the critical point the transition is of second order.
We make a detailed investigation of the critical properties of the
electroweak theory at the critical point, and we find that the
transition falls into the 3d Ising universality class.  The continuum
limit extrapolation of the critical Higgs mass is $m_{H,c} =
72(2)$\,GeV, which implies that there is no electroweak phase
transition in the Standard Model.

\end{abstract}

\vskip1.5pc]

\narrowtext

\footnotetext{$^*$Presented by K. Rummukainen
at the 5th International Workshop on Thermal Field Theories 
and their Applications, Regensburg, Germany, August 1998}


\section{Introduction}

During the last several years, an impressive amount of quantitative
knowledge about the Standard Model (SM) finite temperature
electroweak phase transition has been obtained through lattice Monte
Carlo simulations.  At small Higgs masses ($m_H$) the transition is of
the first order.  The transition becomes weaker (decreasing latent
heat, surface tension) when the Higgs mass increases, and it has been
found to turn into a regular cross-over when $m_H \gsim 75$\,GeV\@
\cite{isthere,karsch,gurtler,4d}.  At the endpoint of the first order
transition line a {\em second order\,} phase transition appears.  At
this point the macroscopic behaviour of the system is determined by
the {\em universal\,} properties of the endpoint.  While the location
of the endpoint and the mass spectrum near it have been studied
before, the critical properties of the endpoint itself have not been
resolved so far.  We shall show that the universality class of the SM
endpoint is of the 3d Ising type (for a full description of this work,
see ref.~\cite{endpoint}).

Near the critical point the thermodynamics of the system
(susceptibilities, correlation lengths) is determined by the
corresponding {\em critical exponents}, which, in turn, are determined
by the universality class of the theory.  Thus, the universal
behaviour represents a tremendous simplification in the effective
degrees of freedom of the system.

While the standard perturbative analysis can be used to
resolve the first order nature of the transition at small $m_H$, it
fails completely at large values of $m_H$ --- indeed, according to 
perturbation theory, the transition remains of first order for all
Higgs masses.  Thus, the physics of the endpoint and the universal
behaviour are inherently non-perturbative.

What kind of universal behaviour can one expect?  Formally, the Higgs
field has ${\rm SU}(2)_{\rm gauge}\times {\rm SU}(2)_{\rm custodial}$
symmetry (in the SU(2) + Higgs theory, see below), but this remains
unbroken at all temperatures.  Indeed, the mass spectrum of the system
has been investigated in detail both above and below the critical
point, and only one scalar excitation (which couples to
$\phi^\dagger\phi$) becomes light in the neighbourhood of it.  Thus,
we expect the Ising-type universality to be realized, but also mean
field-type or multicritical behaviour is, in principle, possible.

\section{Effective action}

An effective 3d SU(2) gauge + Higgs theory, obtained
through {\em dimensional reduction\,}, accurately describes 
the static properties of the SM and many 
of its extensions at high temperatures \cite{generic}.  The
action of the theory is
\be
 S = \int d^3 x
 \bigg[ \fr14 F^a_{ij}F^a_{ij}+|D_i\phi|^2 +
               m_3^2|\phi|^2+
               \lambda_3 |\phi|^4 \bigg].
\label{action}
\ee
The physics of the theory is fixed by the dimensionful gauge coupling
$g_3^2$ and by the dimensionless ratios
\be
 x=\lambda_3/g_3^2, \h y=m_3^2(\mu)/g_3^4\,,
\ee
where $m_3^2(\mu)$ is the renormalized mass parameter in the
$\msbar$ scheme.  The relations of the couplings $g_3^2,x,y$ to
the full theory are computable in perturbation theory~\cite{generic}.
We have omitted the U(1) sector of the SM; this is justified,
since the U(1) gauge boson remains massless at any temperature and
does not affect the transition qualitatively~\cite{su2u1}.


The lattice action in standard formalism is
\begin{eqnarray}
S&=& \beta_G \sum_{x;i,j} (1-\half \tr P_{ij}) \nonumber \\
 &-& \beta_H \sum_{x;i}
\half\tr\Phi^\dagger(x)U_i(x)\Phi(x+i)
\nonumber \\
 &+& \sum_x \left[
\half\tr\Phi^\dagger\Phi + \beta_R
 \bigl( \half\tr\Phi^\dagger\Phi-1 \bigr)^2 \right]
\la{latticeaction} \\
&\equiv& S_G+S_{\rm hopping}+S_{\phi^2}+S_{(\phi^2-1)^2}.
\nonumber
\end{eqnarray}
Here $\Phi$ is the $2\times2$ matrix $\Phi=(i\sigma_2\phi^*,
\phi)$. The two actions in \eqs\nr{action}, \nr{latticeaction} give
the same physics in the continuum limit $a \rightarrow 0$ if the three
dimensionless parameters $\beta_G, \beta_H,\beta_R$ in
\eq\nr{latticeaction} are related to the three dimensionless
parameters $g_3^2a,x,y$ in \eq\nr{action} by the following
equations~\cite{contlatt}:
\begin{eqnarray}
\beta_G & = & {4\over g_3^2a}, \la{betag}\\
\beta_R & = & {\beta_H^2\over\beta_G}x, \la{betar}\\
y & = &
{\beta_G^2\over8}\biggl({1\over\beta_H}-3-
{2x\beta_H\over\beta_G}\biggr)+{3\Sigma\beta_G\over32\pi}
(1+4x)
\nonumber\\
&+& {1\over16\pi^2}\biggl[\biggl({51\over16}+9x-12x^2\biggr)
\biggl(\ln{3\beta_G\over2}+\zeta\biggr) \nonumber\\
&+& 4.9941 +5.2153 x\biggr],
\la{y}
\end{eqnarray}
where $\Sigma=3.1759115$ and $\zeta=0.08849(1)$.
The universal behaviour does not depend on the lattice spacing,
which we keep fixed through fixing $\beta_G = 5$.

\begin{figure}[tb]

\vspace{2mm}
\epsfxsize=7.7cm\epsfbox{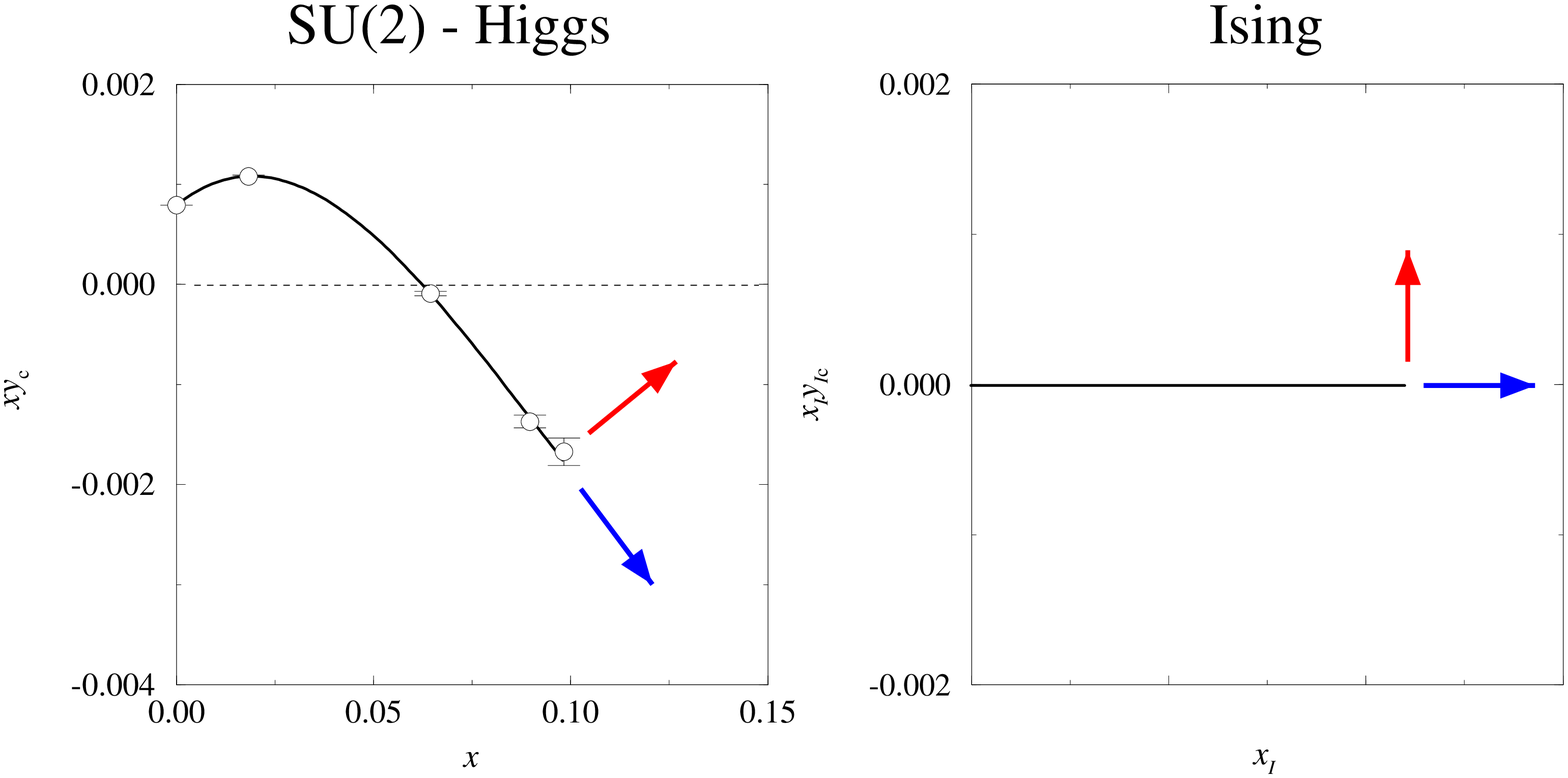}

\vspace*{-1.5cm}
\caption[a]{
The phase diagrams of the SU(2)+Higgs (left) and
the Ising (right) models.
}
\la{fig:phasediag}
\end{figure}


\section{Resolving the universality class}
\label{sec:univ}

For concreteness, in what follows we shall discuss the universal
properties of the SM by comparing it to the Ising model.  However,
one should bear in mind that the analysis is, by no means, limited
only to verifying the Ising-type universal behaviour.
  
The phase diagram of the 3d SU(2)+Higgs theory is shown in
\fig\ref{fig:phasediag}, together with the Ising model phase diagram.
Adopting now Ising-model terminology, let us call the two critical
directions in \fig\ref{fig:phasediag} the $h$-like (perpendicular to
the transition line) and the $t$-like (along the transition line)
directions.  Due to the lack of an exact order parameter, the mapping
of the $h$-like and $t$-like directions of the SU(2)+Higgs model to
the Ising model is non-trivial.  This is illustrated in
\fig\ref{fig:density}, where the probability density at the critical
point is plotted on the $(S_{\rm hopping},S_{(\phi^2-1)^2})$-plane.
Only after a suitable rotation of the axes is the striking similarity
with the Ising model revealed.

\begin{figure}[tb]

\vspace{4mm}
\centerline{\epsfxsize=5cm \epsffile{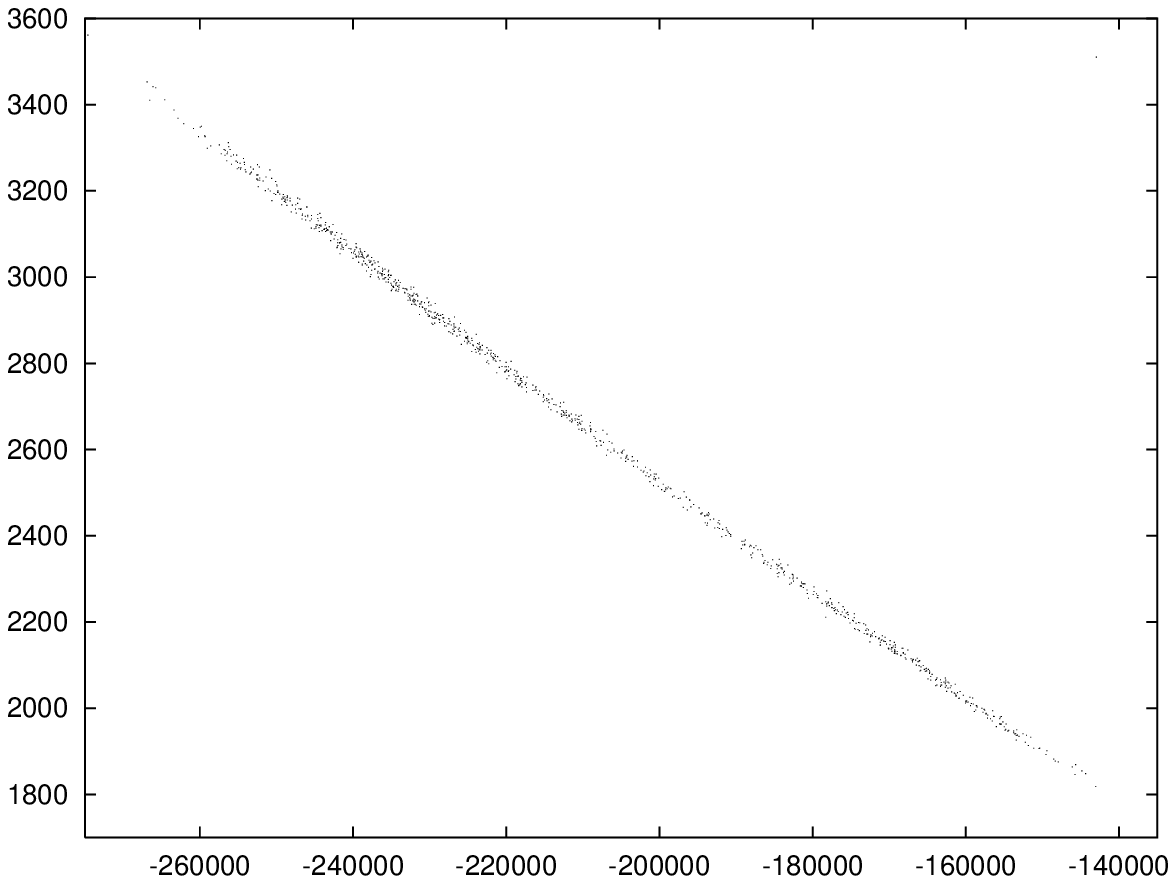}}

\centerline{\epsfxsize=5cm \epsffile{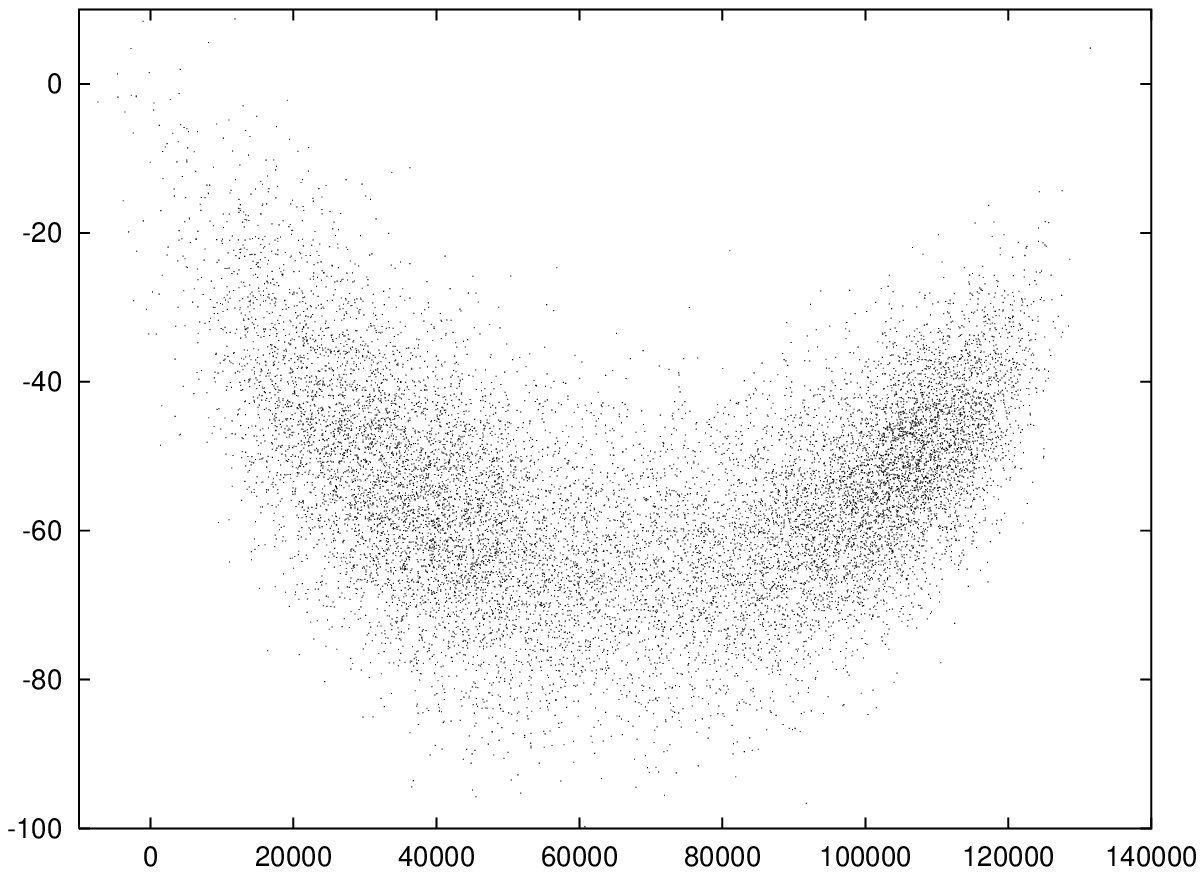}}

\centerline{\epsfxsize=5cm \epsffile{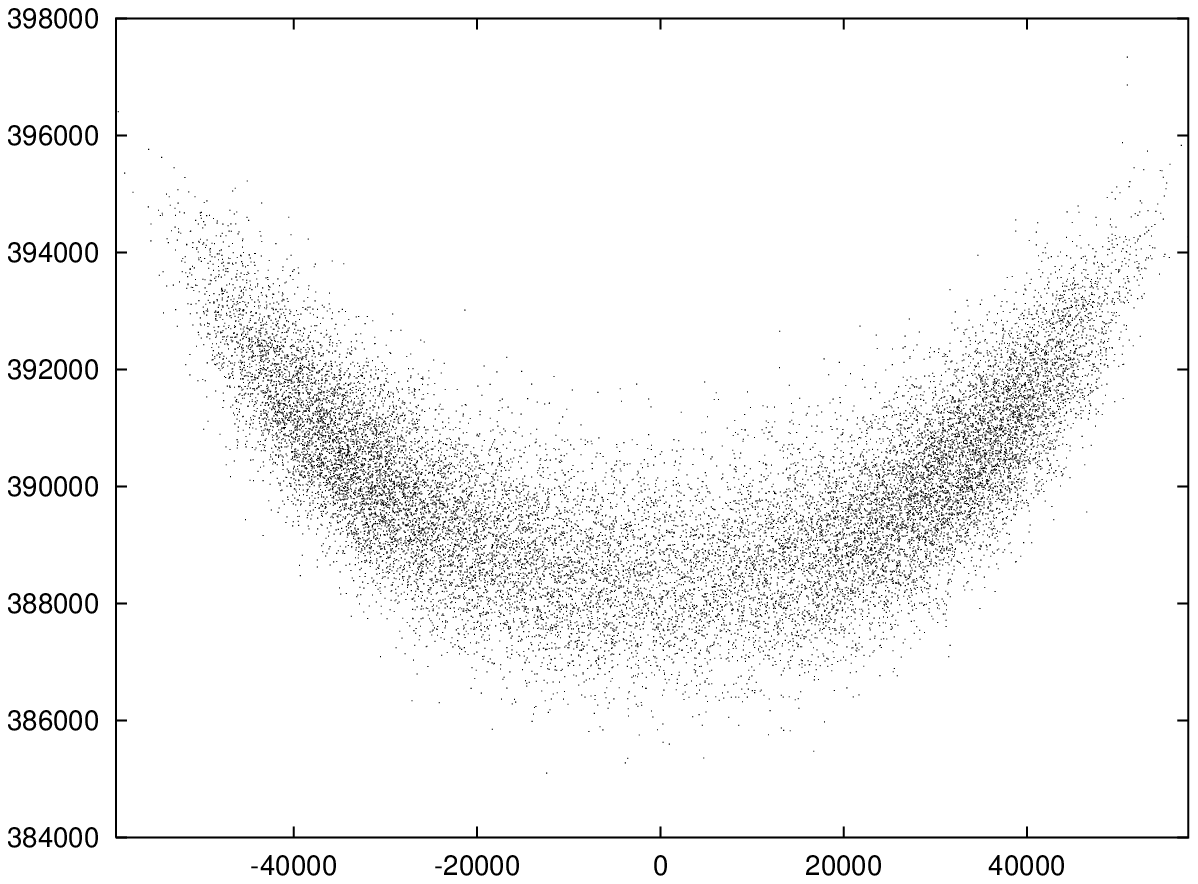}}

\vspace{3mm}
\caption[a]{ {\em Top:} A density plot of the 3d SU(2)+Higgs model at
the critical point, shown on the $S_{(\phi^2-1)^2}$ vs.\ $S_{\rm
hopping}$ plane.  Each point in the plot corresponds to one measurement
of the observables. {\em Middle:} The same as above, after a shift and
a rotation.  {\em Bottom:} A density plot of the
3d Ising model on the $(-1\times\mbox{energy})$ vs. magnetization plane.\la{fig:density}}
\end{figure}


\begin{figure}[tb]

\vspace{4mm}
\centerline{
(a)
\epsfysize=5.0cm \epsfbox[36 40 539 468]{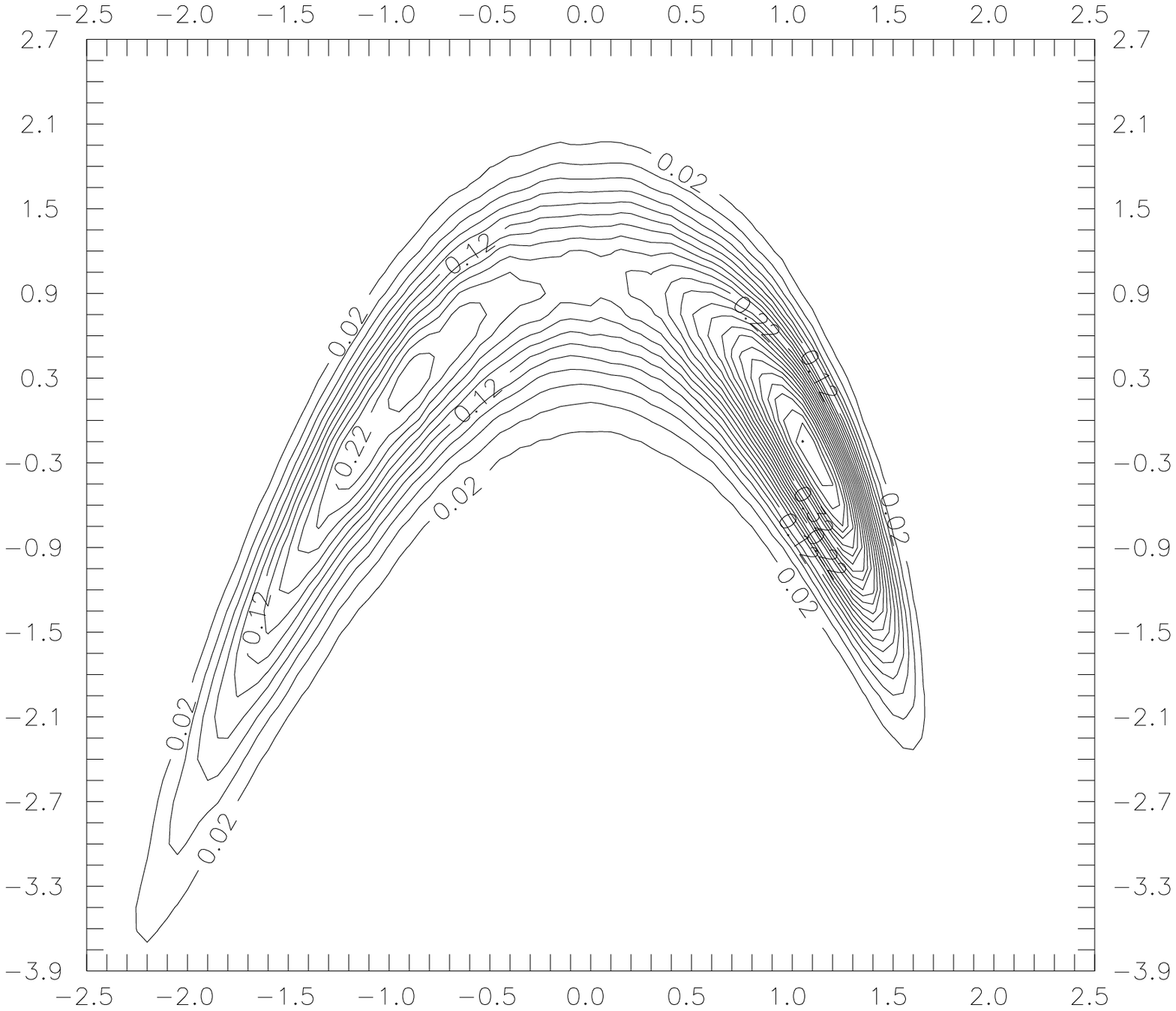}
}
\centerline{
(b)
\epsfysize=5.0cm \epsfbox[36 40 539 468]{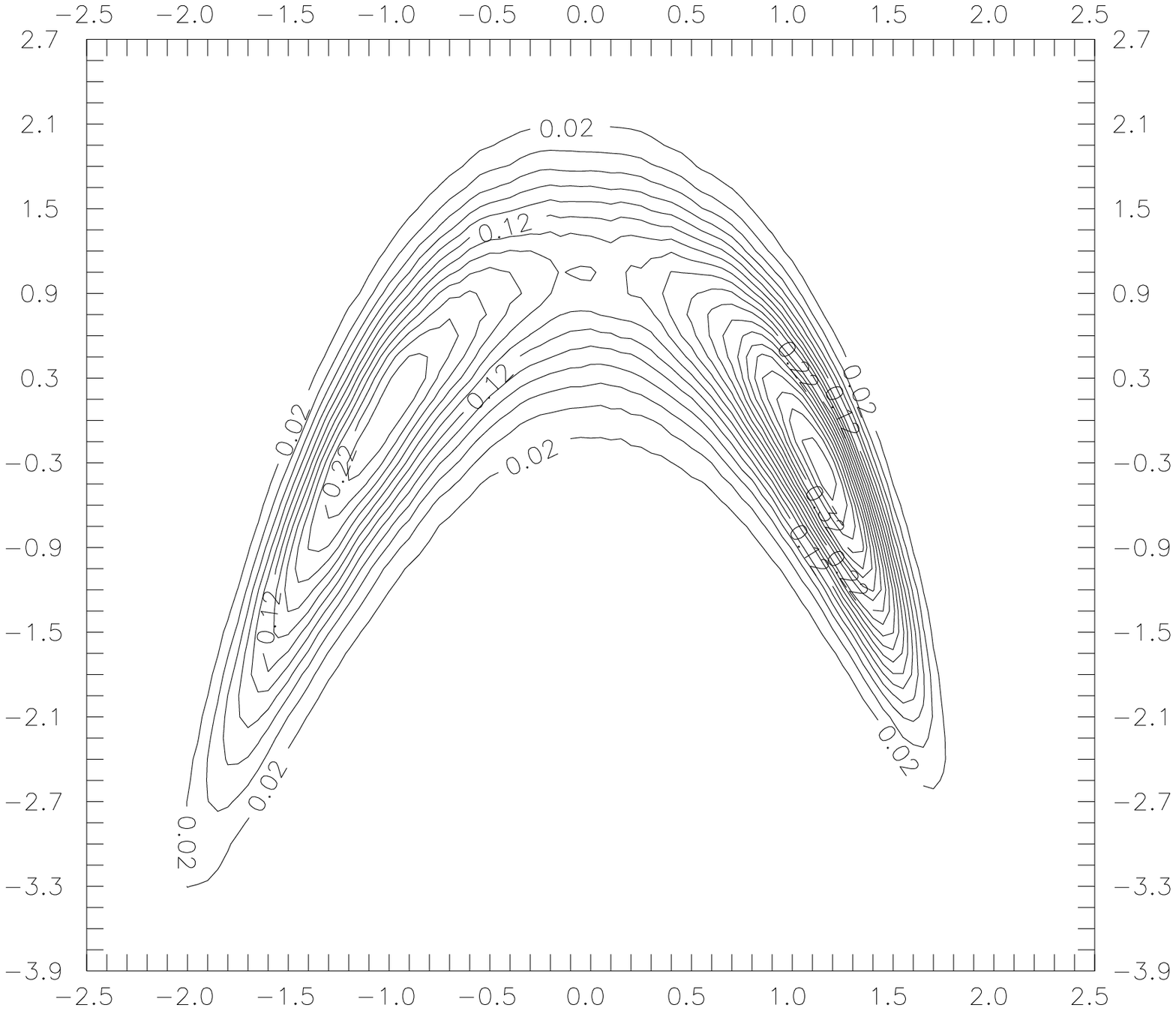}
}
\centerline{
(c)
\epsfysize=5.0cm \epsfbox[36 40 539 468]{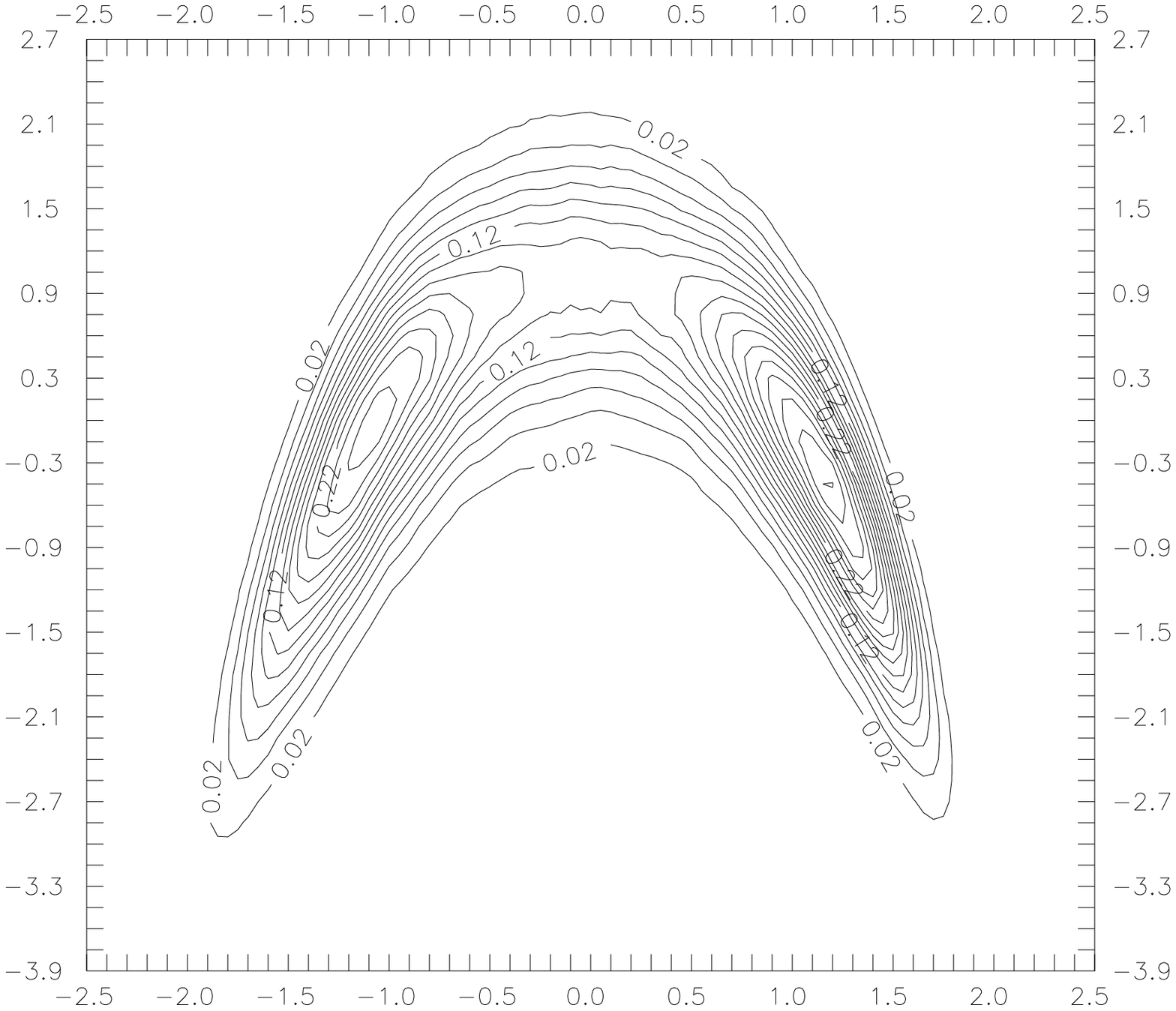}
}
\vspace{4mm}
\caption[a]{The joint probability distributions $P(V_M,V_E)$ at the
infinite volume critical point, for the volumes (a) $16^3$, (b)
$32^3$, (c) $64^3$. It is clearly seen how the distribution becomes
more symmetric for the increasing volume, and approaches the Ising
model distribution of \fig\ref{fig:ising}.} \la{volumes}
\end{figure}

This rotation closely corresponds to the rotation of the directions
shown in the phase diagrams in \fig\ref{fig:phasediag}.  The resulting
rotated operators are conjugate to the $h$-like and $t$-like
couplings, and we call them the $M$-like (``magnetization'') and
$E$-like (``energy'') operators.  Note that all of these operators
($S_{\rm hopping}$, $S_{(\phi^2-1)^2}$ and the rotated $M$-like
operator) lack the explicit $M \leftrightarrow -M$ symmetry present in
the Ising model.  This symmetry is dynamically generated at the
critical point.

However, there is no reason to restrict ourselves only to the two
observables $S_{\rm hopping}$ and $S_{(\phi^2-1)^2}$.  Any number of
(scalar) operators can contribute to the true $M$-like and $E$-like
directions.  In order to improve on the projection, it is important to
consider a large number of independent operators.  The central problem
of the analysis is how to correctly identify the best $E$-like and
$M$-like projections.

Motivated by these considerations, we use the following
method:

\begin{itemize}
\item[(a)] We first locate the infinite volume critical point (for details,
see ref.~\cite{endpoint}), where all of the subsequent analysis
is performed.

\item[(b)] Using several volumes, we measure the {\em fluctuation matrix\,}
$M_{ij} = \langle s_i s_j \rangle$, $ s_i \equiv S_i - \langle
S_i\rangle$.  We used up to 6 operators: those in
\eq\nr{latticeaction}, together with the operators
\ba
S_R&=&\sum_x |\Phi|                  \la{sr} \\
S_L&=&\sum_{x,i} \half\tr V^\dagger(x)U_i(x)V(x+i), \la{sl}
\ea
where $V(x) = \Phi(x)/|\Phi(x)|$.

\item[(c)] We calculate the eigenvalues $\lambda_\alpha$ and eigenvectors
$V_\alpha$ of $M_{ij}$.  Some of the eigenvectors correspond to
``critical'' observables like $M$ or $E$, and the rest are
``trivial.''  They can be classified either by inspecting the
probability distributions $p(V_\alpha)$ and $p(V_\alpha,V_\beta)$, or
by looking at the finite volume behaviour of the eigenvalues.  It is
convenient to express the eigenvalues in terms of the corresponding
susceptibilities: $\chi_i \equiv \lambda_i/L^3$.  For example, the
$M$-like and $E$-like susceptibilities (magnetic susceptibility and
heat capacity) diverge with the critical exponents as ($L$ is the
linear extent of the lattice)
\be
  \lambda_M  \propto L^{\gamma/\nu}\,, \h 
  \lambda_E  \propto L^{\alpha/\nu}\, .   
\ee
The susceptibilities corresponding to the ``trivial'' eigenvalues
remain constant.

\end{itemize}
The method described above has much in common with the one used by
Alonso \etal~\cite{alonso} to find the $E$- and $M$-like directions at
the critical point of the 4d U(1)+Higgs model, and with the method
developed by Bruce and Wilding~\cite{wilding} for the study of the
liquid-gas critical point.  Both of these rely on considering
only two-dimensional distributions.

\begin{figure}[tb]

\vspace{4mm}
\centerline{
\epsfysize=5.0cm \epsfbox[36 40 539 468]{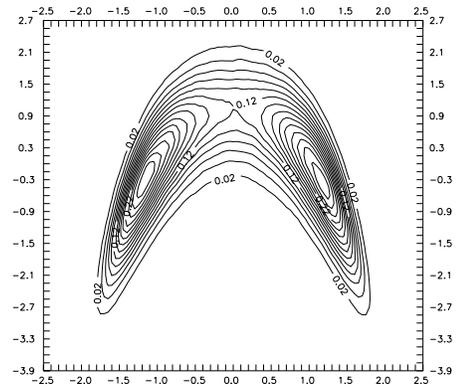}
}
\vspace{4mm}
\caption[a]{The probability distribution $p(M,E)$ for
the Ising model at the critical point (volume $58^3$).}
\la{fig:ising}
\end{figure}

\section{Results}

The main part of our analysis was done on 6 lattice volumes, from
$16^3$ up to $64^3$, using the lattice spacing $a \equiv 
4/(g_3^2\beta_G) = 4/5g_3^2$.  All of the simulations were performed
close to the critical point, and the measurements were
shifted to the measured infinite volume critical point by histogram
reweighting~\cite{endpoint}.  

\begin{figure}[t]

\vspace{2mm}
\hspace*{5mm}\epsfxsize=7cm\epsfbox{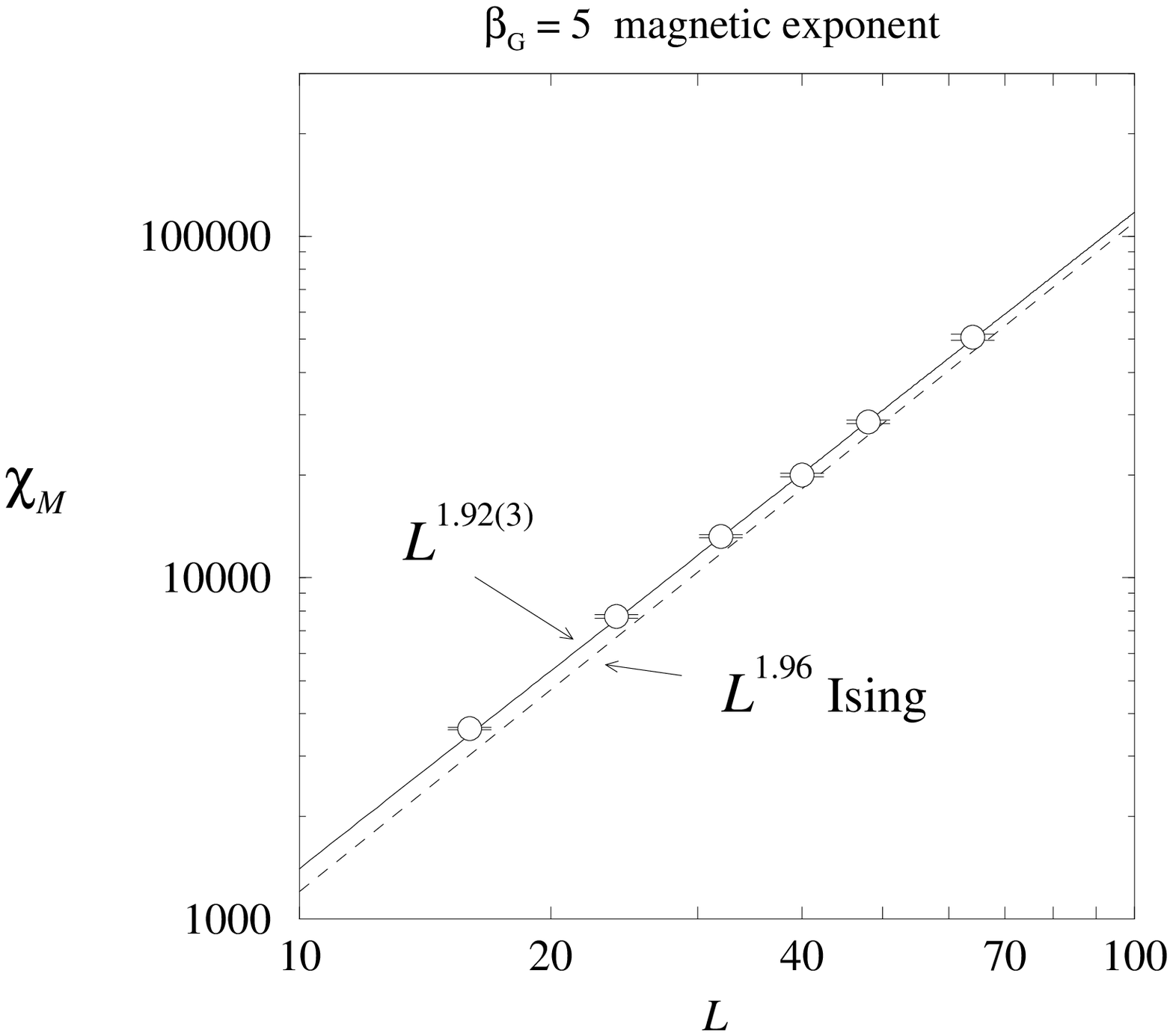}

\hspace*{5mm}\hspace{6mm}\epsfxsize=6.5cm\epsfbox{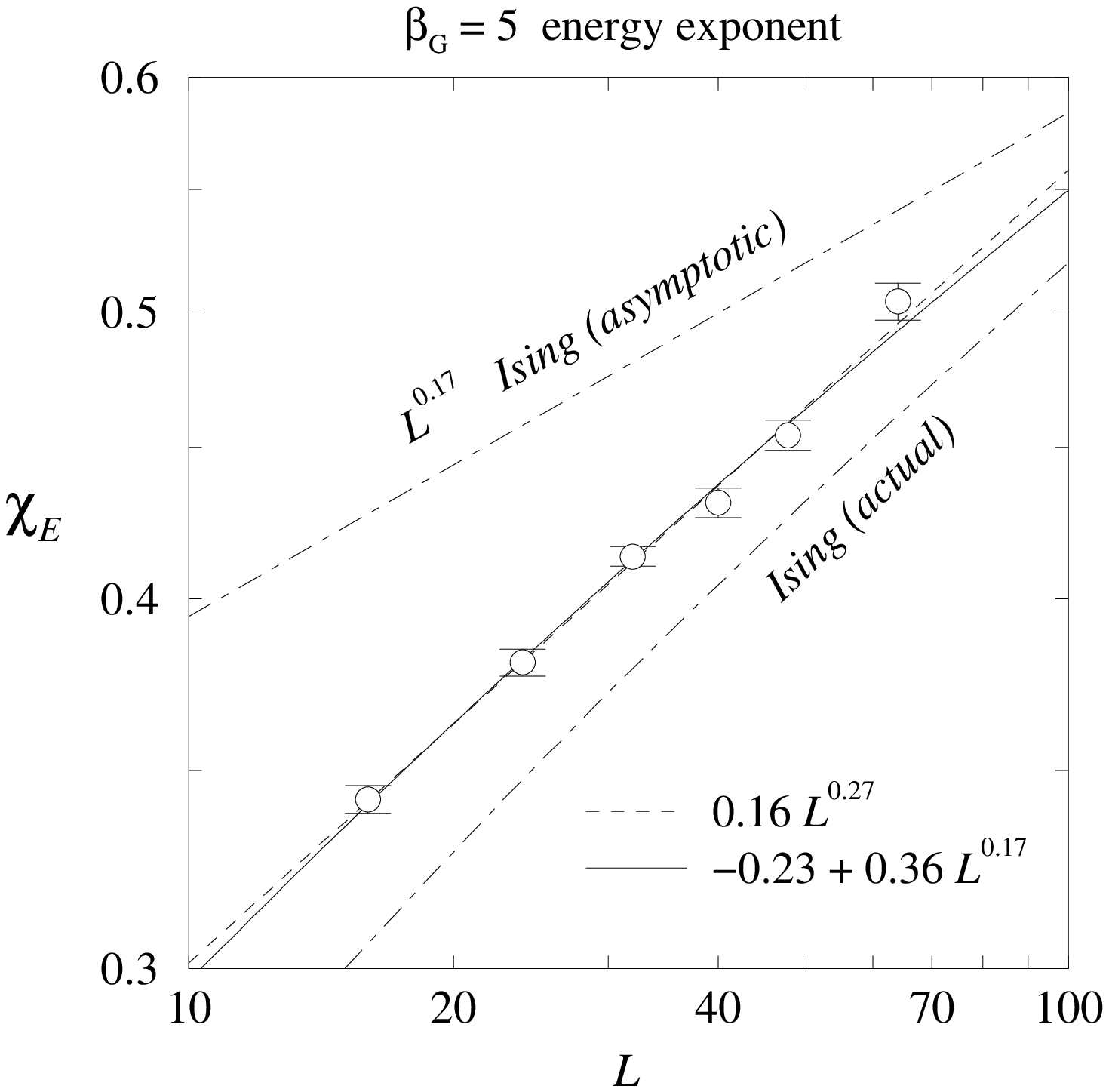}

\vspace{2mm}
\caption[a]{
The divergence of $\chi_M$ ({\em top})
and $\chi_E$ ({\em bottom}) as a function of the lattice size.
The slope of $\chi_E$ does not agree with the
asymptotic Ising value, but it is 
quite consistent with the {\em measured\,} \cite{3dising}
Ising model behaviour at these lattice sizes.}
\la{fig:chis}
\end{figure}

The fluctuation matrix analysis described in the previous section was
performed either using all of the 6 terms described in the item (b)
above or only the 4 terms in the action \nr{latticeaction}.  In both
cases the projections of the eigenvectors to the original operators
remained remarkably stable as functions of the volume.  This gives us
confidence that the method can identify the correct critical observables
and their volume dependence.

It is very illustrative to consider the joint probability
distributions $p(V_M,V_E)$.  In \fig\ref{volumes} these are shown for
volumes $16^3$, $32^3$ and $64^3$ (using 6 operators).  When the
volume increases $p(V_M,V_E)$ clearly approaches the Ising model
distribution $p(M,E)$, shown in \fig\ref{fig:ising}.  It is also
evident that even when the SU(2)+Higgs model lacks the exact
magnetization symmetry of the Ising model, this symmetry is recovered
in the infinite volume limit.  In fact, the shapes of the
distributions $p(V_i,V_j)$ can be used to identify the correct
$E$-like and $M$-like eigenmodes.

\begin{figure}[t]

\vspace{2mm}
\hspace{10mm}\epsfxsize=6.5cm\epsfbox{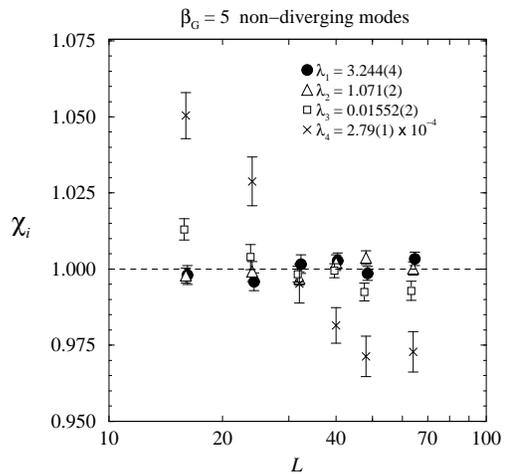}

\vspace{2mm}
\caption[a]{As in \fig\ref{fig:chis}, but for the
regular observables.  The normalization is chosen so that
the average of the values is 1.0; the natural scale of the
eigenvalues is shown in the legend.}
\la{fig:regular}
\end{figure}

In \figs\ref{fig:chis} and \ref{fig:regular} we show the behaviour of
the susceptibilities $\chi_i$ as functions of the volume.  Only
$\chi_M$ and $\chi_E$ diverge as the volume is increased, the other 4
susceptibilities do not show any critical behaviour (note the very
restricted vertical scale in \fig\ref{fig:regular} in comparison with
\fig\ref{fig:chis}).  The behaviour of the
critical susceptibilities $\chi_M$ and $\chi_E$ is compatible
with the Ising model (with the reservation that $\chi_E$ does not
yet attain its asymptotic large volume behaviour; however, it is very
close to the Ising model behaviour at similar lattice sizes).

The divergence of $\chi_E$ implies a positive value for the critical
exponent $\alpha$.  This clearly excludes O($N$) models with
$N\ge 2$ ($\alpha < 0$) and the mean field behaviour ($\alpha = 0$).

The results presented above were calculated using 6 operators in the
fluctuation matrix analysis.  The numerical results remain stable when
only 4 operators are used, although in some cases small deviations
from the Ising model behaviour begin to appear: for example, the joint
distributions $p(V_M,V_E)$ in \fig\ref{volumes} become slightly
thicker to the $E$-direction.  This is in itself not surprising, since
in the case of 4 observables the $E$-like eigenmode has the {\em
smallest} eigenvalue, and presumably it is very sensitive to the
quality of the projection to the operator basis~\cite{endpoint}.  When
the number of the operators is increased to 6, the two additional
eigenvalues are smaller than the $E$-like one.  This underlines the
importance of including a large enough number of operators in the
analysis.

\begin{figure}[t]

\vspace{2mm}
\hspace*{10mm}\epsfxsize=6.5cm\epsfbox{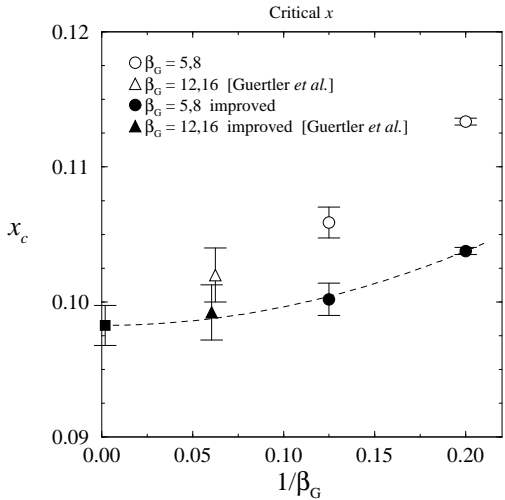}

\vspace{2mm}
\caption[a]{ The infinite volume extrapolations of $x_c$ as a function
of $\beta_G$.  G\"urtler et al refers to~\cite{gurtler}.  The
$O(a)$ -correction has been calculated in ref.~\cite{moore}.}
\la{xcrit_betag}
\end{figure}

In the simulations above the lattice spacing was fixed through
$\beta_G = 4/g_3^2a = 5$.  We have also located the critical point at
$\beta_G = 8$, and G\"urtler \etal~\cite{gurtler} have published
results at $\beta_G = 12$ and 16.  Taking into account the
$O(a)$-correction calculated by Moore~\cite{moore}, we can extrapolate
the critical coupling $x_c(\beta_G)$ to the continuum limit
$\beta_G\rightarrow\infty$ (see \fig\ref{xcrit_betag}).  The final
result $x_c = 0.0983(15)$ corresponds to the Standard Model Higgs mass
$72(2)$\,GeV\@.  The effect of the U(1) gauge sector of the SM, which
was omitted here, is much smaller than the statistical errors
quoted here~\cite{newu1su2}.  Since the experimental lower limit is $\sim
88$\,GeV \cite{mHlower}, this excludes the existence of the SM phase
transition.  Nevertheless, a first order phase transition is still
allowed in several extensions of the Standard Model, including the
Minimal Supersymmetric Standard Model.

\end{document}